\documentclass[trackchanges, twocolumn]{aastex701}

\usepackage[nohyperlinks,nolist]{acronym}

\newcommand{\cmcObsCircSNREccMedian}{0.03}

\newcommand{\cmcObsCircSNREccUpperBound}{0.3}
\newcommand{\cmcObsCircSNRSnrMedian}{13}
\newcommand{\cmcObsCircSNRSnrLowerBound}{10}
\newcommand{\cmcObsCircSNRSnrUpperBound}{28}
\newcommand{\cmcObsEccSNREccMedian}{0.3}
\newcommand{\cmcObsEccSNREccLowerBound}{0.1}
\newcommand{\cmcObsEccSNREccUpperBound}{0.5}
\newcommand{\cmcObsEccSNRSnrMedian}{20}
\newcommand{\cmcObsEccSNRSnrLowerBound}{13}
\newcommand{\cmcObsEccSNRSnrUpperBound}{57}

\newcommand{\cmcObsCircSNRSnrMedianNoise}{12}

\newcommand{\cmcObsEccSNREccMedianNoise}{0.3}

\newcommand{\cmcObsEccSNRSnrMedianNoise}{18}

\begin{document}

\begin{acronym}
  \acro{GW}[GW]{gravitational wave}
  \acro{SNR}[SNR]{signal-to-noise ratio}
  \acro{LVK}[LVK]{LIGO-Virgo-KAGRA}
  \acro{BH}[BH]{black hole}
  \acro{NS}[NS]{neutron star}
  \acro{BNS}[BNS]{binary neutron star}
  \acro{BBH}[BBH]{binary black hole}
  \acro{NSBH}[NSBH]{neutron star--black hole}
  \acro{CMC}[CMC]{Cluster Monte Carlo}
  \acro{O4}{fourth \ac{LVK} Observing Run}
  \acro{GWTC-4}[GWTC-4]{fourth Gravitational Wave Transient Catalog}
  \acro{GWTC-3}[GWTC-3]{third Gravitational Wave Transient Catalog}
  \acro{GC}[GC]{globular cluster}
  \acro{NSC}[NSC]{nuclear star cluster}
  \acro{AGN}[AGN]{active galactic nuclei}
  \acro{IMR}[IMR]{inspiral-merger-ringdown}
  \acro{EOB}[EOB]{effective one body}
  \acro{PN}[PN]{post-Newtonian}
  \acro{NR}[NR]{numerical relativity}
  \acro{FDR}[FDR]{false discovery rate}
  \acro{FPR}[FPR]{false positive rate}
\end{acronym}

\title{Observability of eccentricity in a population of merging compact binaries}

\author[orcid=0000-0001-8081-4888,gname=Mukesh, sname=Singh]{Mukesh Kumar Singh}
\affiliation{Gravity Exploration Institute, School of Physics and Astronomy, Cardiff University, Cardiff CF24 3AA, United Kingdom}
\email[show]{SinghM37@cardiff.ac.uk}  

\author[orcid=0009-0002-2599-1288]{Ben G. Patterson}
\affiliation{Gravity Exploration Institute, School of Physics and Astronomy, Cardiff University, Cardiff CF24 3AA, United Kingdom}
\email[show]{PattersonB1@cardiff.ac.uk}  

\author[orcid=0000-0001-8480-1961]{Stephen Fairhurst}
\affiliation{Gravity Exploration Institute, School of Physics and Astronomy, Cardiff University, Cardiff CF24 3AA, United Kingdom}
\email{}


\begin{abstract}
We investigate the prospects of observing residual eccentricity in a population of compact binaries by calculating the power in the eccentric harmonics, following the methodology in \citet{Patterson:2024vbo}.  Although most observed compact binary coalescences are expected to circularize before entering the sensitivity band of the ground-based \ac{GW} detectors, dynamical interactions in dense star clusters can lead to a fraction of these binaries with non-negligible eccentricity at the time of detection.  To quantify the observability of eccentricity, we simulate a population of merging compact binaries and identify those which have sufficient power in sub-dominant eccentric harmonics to be clearly distinguishable from quasi-circular systems. We consider a \ac{BBH} population derived from globular cluster simulations with residual eccentricity distribution obtained from \ac{CMC} catalogs as well as a fiducial log-uniform model.  Assuming the LIGO-Virgo network of \ac{GW} detectors with their sensitivities achieved during \ac{O4}, we find that the \ac{BBH} population with measurable eccentricity will have a significantly higher median eccentricity $e_{\mathrm{10Hz}}\sim \cmcObsEccSNREccMedian$ (with $90\%$ range: $\cmcObsEccSNREccLowerBound - \cmcObsEccSNREccUpperBound$) and \ac{SNR} $\sim \cmcObsEccSNRSnrMedian$ ($90\%$ range: $\cmcObsEccSNRSnrLowerBound - \cmcObsEccSNRSnrUpperBound$) compared to the observable population of \acp{BBH}. We compare our predictions of the regions of parameter space where eccentricity is detectable with the claimed observations of eccentricity in \ac{GW} events from \ac{GWTC-3}.

\end{abstract}



\section{Introduction}
\label{sec:intro}

In the \ac{GWTC-4}, the \ac{LVK} collaboration announced the observation of over 200 confident \ac{GW} signals from the mergers of compact binaries \citep{GWTC-4}. These discoveries have revolutionized our understanding of the properties of stellar mass \acp{BH} and, to a lesser degree, \acp{NS}. In particular, the measurement of masses and spins of the component \acp{BH} has shed new light on their formation processes \citep{GWTC-1-populations, GWTC-2-populations, GWTC-3-populations, GWTC-4-populations}. The rapid growth in the number of \ac{GW} detections has shifted our scientific emphasis from individual event properties to their population-level inferences \citep{GWTC-4-populations}.  However, the further observations are required to fully understand the formation and evolutionary pathways of compact binary mergers. The proposed formation channels for compact binary mergers can broadly be categorized into two pathways: (i) isolated field evolution, where massive stars in a binary evolve together via transfer of matter, undergo supernovae, and eventually form a compact binary of \acp{BH} or \acp{NS}, or both \citep{Bethe:1998bn, PortegiesZwart:1997ugk, Belczynski:2001uc, Hurley:2002, Dominik:2014yma, Mandel:2015qlu, Marchant:2016wow, deMink:2016vkw, Neijssel:2019irh, vanSon:2022sma}; and (ii) dynamical formation, involving gravitational encounters from compact objects in dense stellar environments, such as, \acp{GC} \citep{Kulkarni:1993fr, Sigurdsson:1993gcs, PortegiesZwart:1999nm, Ziosi:2014sra}, \acp{NSC}, or the disks of \acp{AGN} \citep{Mckernan:2017ssq, Bartos:2016dgn, Stone:2017agn, Fragione:2018yrb}. Current \ac{GW} observations suggest that no single formation channel is sufficient to explain the observed population of \ac{BBH} mergers; rather, a mixture of these two pathways likely contributes to the observed distribution \citep{Zevin:2020gbd}.

Although most binaries shed all traces of their evolutionary history when they enter the frequency band of ground-based \ac{GW} detectors, some of the features observed through \ac{GW}  signals can still give clues to their formation and evolution pathways \citep{Mandel:2018hfr, Mapelli:2018uds}. For example, isolated field evolution typically leads to spins of the component objects aligned with the orbital angular momentum in a binary due to the tidal and matter interactions between them \citep{Kalogera:2000som, Gerosa:2018wbw, Steinle:2022rhj, Wysocki:2018nk, Stevenson:2022hmi}. While dynamical interactions can also lead to aligned spins \citep{Rodriguez:2015gcs, Rodriguez:2016kxx,Rodriguez:2018dsl, Farr:2017uvj}, both misaligned spins \citep{Trani:2021som, Banerjee:2023ycw, Kiroglu:2025bbp} and non-circular orbits \citep{Samsing:2013kua, Samsing:2017xmd, Rodriguez:2017pec} can occur primarily due to dynamical interactions, making them unique signatures to constrain the formation of compact binaries. In practice, it is often challenging to robustly measure the misaligned component of the spins \citep{GWTC-2:2020ibl, GWTC-3:2021vkt, Fairhurst:2019srr, Vitale:2014ms}. Therefore, orbital eccentricity becomes a unique and robust tracer of the dynamical origin of a compact binary merger \citep{Samsing:2013kua, Samsing:2017oij, Samsing:2017xmd, Rodriguez:2017pec}. The binaries formed in isolation generally have quasi-circular orbits \citep{Belczynski:2001uc}~\footnote{It has been pointed out that large supernova natal kicks, asymmetric common envelope ejection can lead to binaries with residual eccentricity in the field evolution channel \citep{Brandt:1995ske, Kalogera:2000som, Hurley:2002, Ivanova:2013ce}.}, while in contrast, gravitational interactions such as three body encounters or secular evolution mechanisms such as the Lidov-Kozai mechanism \citep{Lidov:1962wjn, Kozai:1962zz} within dense star clusters can lead to highly eccentric binaries that retain significant eccentricity near the time of merger \citep{Antonini:2012se}.

As the majority of merging compact binaries efficiently radiate any eccentricity away through the emission of \acp{GW} by the time they enter the sensitivity band of ground-based \ac{GW} detectors, it is natural that most \ac{GW} data analysis techniques assume quasi-circular waveform models \citep{Nitz:2017svb, Cannon:2020qnf, Chu:2020pjv, Allene:2025saz, GWTC-2:2020ibl, GWTC-3:2021vkt}. However, as the number of detections grows, the chance of detecting eccentric events increases. Several studies have demonstrated that the assumption of quasi-circular waveform models can lead to loss in the sensitivity of \ac{GW} searches to eccentric signals, potentially causing them to be disregarded as noise \citep{Martel:1999tm, Brown:2009ng, Huerta:2013qb, Lower:2018seu, Romero-Shaw:2019itr, Phukon:2024amh}. Moreover, the parameter estimation of eccentric signals with a quasi-circular waveform model can lead to biases in the inferred properties \citep{Huerta:2013qb, Favata:2013rwa, Lower:2018seu, Romero-Shaw:2019itr, Ramos-Buades:2019uvh, Divyajyoti:2023rht} as well as false violation of general relativity \citep{Narayan:2023vhm, Shaikh:2024wyn}. 

Robust detection and inference of the residual eccentricity in compact binary mergers rely on the availability, fidelity, and computational efficiency of eccentric waveform models. The majority of the existing eccentric waveform models are in the time domain \citep{Islam:2021mha, Liu:2023dgl, Nagar:2024dzj, Gamboa:2024hli, Nee:2025nmh, Paul:2024ujx, Planas:2025feq} and as such are more computationally expensive to evaluate compared to the frequency domain models. In addition, none of the existing eccentric waveform models take into account the full extent of the physics at play in the mergers of compact binaries; for example, there is currently no widely-used waveform model that accurately incorporates both eccentricity and precession, although there have been some efforts toward developing such models \citep{Klein:2018ybm, Klein:2021jtd, Gamba:2024cvy, Morras:2025nlp}. It has been demonstrated that neglecting the degeneracy between these two effects can lead to a biased inference, especially for heavy binary systems \citep{Klein:2018ybm, Romero-Shaw:2020thy, Xu:2022zza,  Divyajyoti:2025cwq}. These factors limit both the speed and accuracy of the parameter estimation of eccentric \ac{GW} signals. While several works have successfully been applied to perform eccentric Bayesian parameter estimation \citep{Bonino:2022hkj, Ramos-Buades:2023yhy, Morras:2025xfu, Planas:2025jny}, the machine learning \citep{Gupte:2024eep} and likelihood reweighting \citep{Romero-Shaw:2019itr} approaches%
\footnote{Recently, \cite{Patterson:2024vbo} has demonstrated that the eccentric harmonics used in this study can, in principle, be applied to perform parameter estimation on eccentric binaries in minutes.}
have been proposed as alternatives to reduce the computational cost. This has enabled the investigation of the presence of residual eccentricity in several \ac{BBH} mergers observed by \ac{LVK}, leading to several claims of signals exhibiting eccentricity \citep{Romero-Shaw:2021ual, Gayathri:2020coq, Gupte:2024eep, Morras:2025xfu, Planas:2025jny}. However, these claims are dependent on the treatment of noise glitches in the data \citep{Gupte:2024eep}, waveform models \citep{Romero-Shaw:2020thy, Gayathri:2020coq}, and most crucially on the choice of priors \citep{Romero-Shaw:2021ual, Romero-Shaw:2025vbc, Morras:2025xfu}.

In this work, we investigate the measurability of residual eccentricity in a population of \ac{BBH} mergers. We exploit the eccentric harmonics decomposition introduced in \cite{Patterson:2024vbo} to compute the power (or \ac{SNR}) in the sub-leading eccentric harmonics which contribute to the measurability of residual orbital eccentricity in the signal. We require the \ac{SNR} in these subdominant eccentric harmonics to be above a reasonably justified threshold in order for the residual eccentricity to be measurable. We explore the observable population of the \acp{BBH} most likely to be detected as eccentric and their corresponding \ac{SNR} in \ac{O4} of \ac{LVK} detector network, and compare this to existing claims of observable eccentricity from \ac{GWTC-3}. 

The paper is structured as follows. Section \ref{subsec:eccentric_wf_models} discusses various waveform models currently in use, and the one used in this work, for detecting and inferring properties of eccentric \ac{BBH} mergers. Section \ref{subsec:ecc_harmonics} details the \citet{Patterson:2024vbo} framework for eccentric harmonics. Section \ref{subsec:ecc_models} describes the population of \acp{BBH} sampled from \ac{CMC} simulations as well as log uniform eccentricity model. Sections \ref{subsec:detection} and \ref{subsec:IdentifyEcc} discuss the criteria for detection of BBHs and measurability of eccentricity respectively. Sections \ref{subsec:obs_ecc_dist} and  \ref{sec:ecc_vs_snr_scatter} quantify the observability of eccentric binaries, including the population of binaries with observable eccentricity. Section \ref{subsec:ecc_candidates} presents the comparison with observed eccentricity candidates from \ac{GWTC-3}. Section \ref{sec:search} investigates potential searches using eccentric harmonics. Finally, Section \ref{sec:summary} provides a concise summary.

\section{Method}

Our approach to determine the observability of eccentric events is summarized in three key steps. Firstly, we generate an astrophysically motivated population of \ac{BBH}s from globular cluster simulations to act as our starting point. We then apply two cuts to this population: one to determine which events would be detected by current \ac{GW} detectors, and another to determine which events have high enough power in subdominant eccentric harmonics to enable the measurement of eccentricity. Mergers that pass both of these cuts should both be observed and measured as eccentric with current techniques. We also explore the use of eccentric harmonics for searching eccentric \ac{GW} signals and compare their performance against existing quasi-circular searches. We begin the discussion of the methods by summarizing available eccentric waveform models and how we decompose them into eccentric harmonics.

\subsection{Gravitational waveforms for eccentric binaries}
\label{subsec:eccentric_wf_models}

Gravitational-wave signals emitted from binaries on eccentric orbits are generally more complex than their quasi-circular counterparts. Unlike quasi-circular binaries, where the \ac{GW} radiation is predominantly emitted at twice the orbital frequency, and both the amplitude and frequency of the signal evolve monotonically, the presence of eccentricity induces additional non-monotonic amplitude and frequency variations as a function of time. The amplitude of the eccentric \ac{GW} signal is modulated, being the largest at periapsis (smallest separation) and the smallest at apoapsis (largest separation). The \ac{GW} waveform at Newtonian order was first approximated for such non-circular binary systems by \citet{Peters:1963ux, Peters:1964zz} and subsequent efforts were made to compute the \ac{PN} terms for the inspiral of the \ac{GW} signal from compact binaries \citep{Junker:1992kle, Blanchet:1989cu, Gopakumar:1997bs, Memmesheimer:2004cv, Arun:2007rg, Arun:2007sg, Arun:2009mc, Mishra:2015bqa, Boetzel:2019nfw, Ebersold:2019kdc, Morras:2025nlp}.
More recently, the complete waveform models for eccentric \ac{GW} signals have been developed using both the analytical inspiral and quasi-circular numerical simulations for the merger part of the signal (any residual eccentricity is expected to have been radiated away by the end of the inspiral) \citep{Hinder:2017sxy, Nagar:2021gss}. At present, there are two main eccentric waveform models in use for searches and parameter estimation of eccentric compact binary mergers: i) \texttt{TEOBResumS-Dali} \citep{Chiaramello:2020ehz, Nagar:2021gss, Nagar:2024dzj} and ii) \texttt{SEOBNRv4EHM, SEOBNRv5EHM} \citep{Ramos-Buades:2021adz, Gamba:2024cvy}, \texttt{SEOBNRE} \citep{Liu:2023dgl}. Both of these models use the \ac{EOB} formalism \citep{Buonanno:1998gg} to model the full \ac{IMR} \ac{GW} waveforms. Other available eccentric waveform models include \texttt{ESIGMAHM}, that combines \ac{PN} theory and self force corrections to describe the inspiral, and uses \ac{NR} based surrogate model to describe the strong field merger regime \citep{Paul:2024ujx}. There is also \texttt{IMRPhenomTEHM}, a time-domain extension to the \texttt{IMRPhenom} waveform family which has recently been applied to the eccentric parameter estimation of a selection of \ac{GWTC-3} events thanks to its much lower computational cost \citep{Planas:2025feq, Planas:2025jny}. Most of these waveform models account for the effect of aligned spins but do not consider the effect of in-plane spins that lead to the precession of the orbit of the binary, except for the latest developments of \texttt{TEOBResumS-Dali} \citep{Gamba:2024cvy}. In this work, we use aligned-spin version of the \texttt{TEOBResumS-Dali} waveform model to characterize the population of observable eccentric binary mergers.

The definition of eccentricity generally differs across various families of gravitational waveform models, and thus care must be taken when comparing the eccentricity estimates derived using different eccentric waveforms. \ac{NR} simulations and waveform models typically define eccentricity based on gauge-dependent quantities, such as binary orbital separations or trajectories, making it a non-unique observable. To address this ambiguity, several studies have made efforts to bring these definitions to an equal footing \citep{Knee:2022hth, Ramos-Buades:2021adz, Bonino:2022hkj, Shaikh:2023ypz, Shaikh:2025tae}. In particular, \citet{Shaikh:2023ypz} has defined the eccentricity, $e_{\mathrm{gw}}$, at a reference orbit-averaged frequency of the quadrupole mode of gravitational radiation at null infinity, a gauge-independent quantity. In this work, however, it is computationally expensive to estimate $e_{\mathrm{gw}}$ for each of the simulations. Hence, we keep the eccentricities reported consistent with the definition in \texttt{TEOBResumS-Dali}. When the eccentricity measurements are used from different waveform models, they are converted to equivalent eccentricities for \texttt{TEOBResumS-Dali} using the mismatch calculations following \cite{Knee:2022hth}.

\subsection{Eccentric Harmonic decomposition}
\label{subsec:ecc_harmonics}

The complex amplitude and frequency evolution of a gravitational wave emitted by a binary on an eccentric orbit can more easily be understood by decomposing the signal into a series of eccentric harmonics, whose frequencies differ by multiples of the radial frequency. Examining the eccentric harmonic structure of a waveform allows us to identify the power from an event in each individual harmonic, and so quantify the power attributed to eccentricity.  In this work, we restrict attention to the dominant ($l=2, m=\pm2$) mode of \texttt{TEOBResumS-Dali} waveforms and decompose into eccentric harmonics using the method first introduced by \citet{Patterson:2024vbo}. A similar method has also been used to create a surrogate model for the low-eccentricity, non-spinning regime \citep{Islam:2025rjl, Islam:2025llx}. We summarize the approach here and detail improvements we have made. 

We begin by generating $n$ eccentric waveforms, denoted $x_{j}$ with identical parameters but equally spaced mean anomaly at a fixed time before merger. This is equivalent to saying that their eccentricity-induced amplitude modulations are equally spaced in time.  The eccentric harmonics $h_{k}$ are obtained by combining these waveforms, with the appropriate weighting factors to extract individual harmonics as
\begin{equation}
\label{eq:ecc_harms}
h_k = \frac{1}{n}\sum_{j=0}^{n-1}e^{\left(2\pi ijk/n\right)} x_j.
\end{equation}
To accurately extract the harmonics, it is important that the waveforms $x_{j}$ all have the same orbit-averaged eccentricity, frequency and phase at the specified start time, with only the mean anomaly differing. This is done using leading order PN evolution equations \citep{Peters:1963ux, Moore:2018kvz} to evolve the eccentricity, frequency and phase in time to their desired values. The use of leading order PN expressions leads to some inaccuracy in the waveforms, so we have implemented a new method to improve accuracy.  Specifically, we first generate a test waveform whose mean anomaly should differ by $2\pi$ from the original.  Any discrepancy between the two waveforms can be attributed to use of the leading-order PN equations, and empirically corrected by comparing these two waveforms. 

The dominant harmonic, $k=0$, is simply an (equally weighted) average of all of the $x_{j}$ waveforms, effectively suppressing the amplitude modulations present in the individual eccentric waveforms. We therefore think of the $k=0$ harmonic as similar (although not exactly equivalent) to a quasi-circular waveform, with higher harmonics ($k\neq0$) representing the eccentric corrections. These harmonics are orthonormalized using Gram-Schmidt orthonormalization to remove the small amount of residual power shared across harmonics. These eccentric harmonics are implemented into the \texttt{simple-pe} python package \citep{Fairhurst:2023idl} which we use in this study. As shown in \citet{Patterson:2024vbo}, the majority of the power in an eccentric waveform is contained in the dominant $k=0$ eccentric harmonic and the sub-leading $k=\pm1$ harmonics.

\subsection{Population Models}
\label{subsec:ecc_models}

The expected distribution of residual eccentricities for \ac{GW} captures in dense stellar environments relies on several underlying assumptions, including the initial mass function \citep{Samsing:2017xmd}, metallicity, and kick velocity distributions \citep{Antonini:2019ulv}. However, state-of-the-art N-body simulations of globular clusters \citep{Kremer:2019iul} have enabled a robust prediction of the residual eccentricity distribution for \ac{GW} captures \citep{Samsing:2013kua, Rodriguez:2017pec}. Both semianalytical and numerical approaches have shown that \ac{GW} encounters in globular clusters can lead to $\approx 10\%$ eccentric \ac{BBH} mergers, approximately half of which have residual eccentricity $\geq 0.1$ at a \ac{GW} frequency of 10Hz, the starting point of the sensitivity band of the LIGO-Virgo-KAGRA detectors \citep{Samsing:2017rat, Samsing:2017oij, Samsing:2017xmd, Rodriguez:2017pec, Zevin:2018kzq, Kremer:2019iul, Antonini:2019ulv}.  

For this work, we simulate a population of eccentric \ac{BBH} mergers obtained from the \ac{CMC} catalog for dense star cluster models \citep{Rodriguez:2017pec, Kremer:2019iul}. The eccentricity values reported in these simulations are defined at the frequency of the harmonic with the highest power \citep{Wen:2002km}. As noted by \cite{Vijaykumar:2024piy}, this generally differs from the eccentricities corresponding to the orbit-averaged frequency of the quadrupole mode --- the convention currently recommended to standardize the definition of eccentricity in \ac{GW} waveform models \citep{Ramos-Buades:2021adz, Shaikh:2023ypz}. For \ac{CMC}  population, we use the eccentricity ($e^{\mathrm{2PN}}$) defined at orbit-average frequencies of the quadrupole mode and assuming evolution with $\mathrm{2PN}$ order equations, following \citet{Vijaykumar:2024piy}, to generate \texttt{TEOBResumS-Dali} waveforms. We find that \texttt{TEOBResumS-Dali} waveforms generated with $e^{\mathrm{2PN}}$ at a reference frequency of $10$Hz approximately lead to $e_{\mathrm{gw}}$ close to the values if it would have been calculated using initial conditions of bound systems.

We also sample a simulation and theory agnostic population of eccentric \acp{BBH} with residual eccentricity distributed log-uniformly and masses according to the power-law plus peak (PLP) model from GWTC-3 observations \citep{GWTC-3-populations}. From \ac{CMC}  simulations, we observe that high mass \acp{BBH} have low eccentricities, at a fixed reference frequency, as compared to low mass ones. This is because of high mass \acp{BBH} having smaller separations at a fixed reference frequency and hence smaller eccentricities. We maintain this correlation in the log-uniform model as well. The \acp{BBH} are distributed uniformly in comoving volume and source frame time.

\subsection{Event detection}
\label{subsec:detection}

Given a simulated population of binary mergers, as described in Section \ref{subsec:ecc_models}, we first wish to identify those events which would be observable in a network of gravitational-wave detectors.  In this study, we take the sensitivity of the detectors to be given by the representative sensitivity of the LIGO and Virgo observatories during the \ac{O4}  run \citep{O4-psds-LVK}.  The sensitivity of searches varies over the parameter space of binary masses and spins \citep{LIGOScientific:2025yae} and is also impacted by data quality. Here we make the simplifying assumption that an event with a network \ac{SNR} of 10 or higher will be observable. This allows us to immediately discard any events with a total network \ac{SNR} below this threshold. Although searches which incorporate the effects of eccentricity have been performed \citep{LIGOScientific:2023lpe, Wang:2025yac}, all current modelled CBC searches%
\footnote{There are also unmodelled searches which do not make assumptions about the form of the signal which could be sensitive to eccentric signals, see for example \citet{Drago:2020kic, Skliris:2020qax}}%
used by the \ac{LVK} are restricted to a quasi-circular parameter space \citep{Nitz:2017svb, Cannon:2020qnf, Chu:2020pjv, Allene:2025saz}. It is therefore natural to place a detection threshold on the ``quasi-circular SNR'', $\rho_{\mathrm{circ}}$, i.e. the maximum \ac{SNR} found in a \ac{GW} signal when matched filtering against only quasi-circular waveforms. 

To estimate $\rho_\mathrm{circ}$, we matched-filter quasi-circular waveforms generated by \texttt{TEOBResumS-Dali} against the eccentric \ac{GW} signal, to find the maximum \ac{SNR}.  For computational efficiency, we only vary the chirp mass of the system, keeping the mass ratio and spins fixed.   In principle, degeneracies also exist between eccentricity and spin or mass ratio parameters, however we find a negligible contribution from allowing these parameters to vary and so keep them fixed for simplicity and reduced computational cost. This process will give an accurate estimate of $\rho_\mathrm{circ}$ as we are able to find the quasi-circular waveform that gives the highest matched filter \ac{SNR} with a given eccentric signal, and thereby mimic the behaviour of the matched filter template search approaches we are attempting to emulate.

An alternative approach is to calculate $\rho_{k=0}$, the \ac{SNR} contained in the dominant $k=0$ eccentric harmonic in the waveform decomposition introduced in Sec.~\ref{subsec:ecc_harmonics}.  This waveform is generated by averaging over the amplitude and frequency  modulations caused by the eccentricity of the orbit.  It can therefore be thought of as an (approximately) quasi-circular waveform.  While this signal corresponds to an eccentric binary, it has been shown \citep{Favata:2021vhw, Patterson:2024vbo} that an eccentric waveform matches well with a circular waveform for a signal with a higher chirp mass. However, as the eccentricity of the orbit decreases due to gravitational-wave emission, the dominant eccentric harmonic is actually equivalent to a circular waveform with a time-varying chirp mass.  Therefore, filtering the $k=0$ eccentric harmonic against an eccentric signal actually leads to an \ac{SNR} greater than found by a quasi-circular waveform, i.e. $\rho_{k=0} \ge \rho_\mathrm{circ}$. 

As discussed above, current searches use quasi-circular templates when searching the parameter space of binary mergers.  The fact that the \ac{SNR} recovered by the $k=0$ eccentric harmonic is higher than that in the best-matched circular waveform leads us to propose a relatively simple extension to existing searches that would increase sensitivity to eccentric signals.  Performing an eccentric search over the full parameter space is complicated by the fact that the templates need to be laid out in two new parameters, eccentricity and mean anomaly, in addition to masses and spins. However, if we restrict to the $k=0$ harmonic, only a new single parameter, the eccentricity, is required to describe the waveform.  This would reduce the size of the template-bank compared to full eccentric searches.  We discuss the improved detection efficiency of the $k=0$ eccentric harmonic search compared to quasi-circular search in the results in Section \ref{sec:search}. 

\subsection{Identification of eccentricity}
\label{subsec:IdentifyEcc}

We wish to determine whether an observed event would be identified as originating from an eccentric orbit.  To do so, the $k \neq 0$ eccentric harmonics can be used to determine whether they contain sufficient power for eccentricity to be observable.  As shown in \citet{Patterson:2024vbo}, the most important subleading eccentric harmonics are $k = \pm 1$, corresponding to the harmonics of frequency higher ($k=1$) and lower ($k=-1$) than the dominant eccentric harmonic by exactly the radial frequency. We expect that any events with high enough eccentricity where the power in $|k|>1$ eccentric harmonics becomes non-negligible should also have sufficient power in the $k=\pm1$ eccentric harmonics alone to be detected as eccentric, and so we disregard any higher harmonics for simplicity. Therefore, we define the ``eccentric harmonic SNR'', $\rho_\mathrm{ecc}$, as the combined power in the $k=1,-1$ harmonics, as described by Sec. V.C of \citet{Patterson:2024vbo}. This exploits the known phase relation between the eccentric harmonics in order to combine the power from both individual harmonics in a phase consistent way. 

In order to classify an event as observably originating from an eccentric binary, we require a threshold of $\rho_\mathrm{ecc} \geq 4$. For a universe where $1\%$ of observed \acp{BBH} are eccentric, roughly $10\%$ of events that pass this condition would actually be non-eccentric (see Section \ref{subsec:NoiseMethods} for details of this calculation). An event which has $\rho_{\mathrm{circ}} \geq 10$ and $\rho_{\mathrm{ecc}} \geq 4$ is deemed to be observable and observably eccentric.  As discussed above, for some signals the power in the $k = 0$ harmonic is not fully captured by a circular waveform, as $\rho_{k=0} > \rho_{\mathrm{circ}}$.  In this case, there is additional power, orthogonal to a circular template, which carries evidence of eccentricity.  In this analysis, we do not account for that power and simply use a threshold on $\rho_{\mathrm{ecc}}$ as our definition of observable eccentricity. Similar to disregarding higher eccentric harmonics, across much of the \ac{BBH} mass space, this effect should only become important at high eccentricities, where we expect the power in $\rho_\mathrm{ecc}$ alone to be high enough to identify eccentricity in all events that are detected.  However, as discussed in Section \ref{sec:search}, the difference between a quasi-circular waveform and the $k=0$ eccentric harmonic becomes more significant at low masses, so this approach may not generalize to \ac{BNS} and \ac{NSBH} systems.

\subsection{Including the effects of noise}
\label{subsec:NoiseMethods}

In a realistic detection scenario, detector noise will impact the observability of eccentricity in several ways.  Most notably, detector noise will impact the observed value of the \ac{SNR}, differing from the expected \ac{SNR} depending on the exact noise realization.  Similarly, the observed \ac{SNR} in the eccentric harmonics will be affected by detector noise and therefore change the set of signals which would be identified as arising from a binary on an eccentric orbit. Realistic detector noise is affected by periods of poor data quality, and non-stationarities (or glitches) \citep{Nuttall:2018xhi, LIGOScientific:2025yae}.  For this intial study, we do not incorporate these effects and we take the noise to be stationary and Gaussian.  In that case the square of the observed \ac{SNR} of each eccentric harmonic is described by a non-central chi-squared distribution with two degrees of freedom and a non-central parameter as the true \ac{SNR} squared (see e.g.~\cite{Mills:2020thr} for details). Here, the two degrees of freedom arise from the amplitude and phase of the \ac{SNR} or, equivalently, the real and imaginary parts. Introducing noise will also shift the best-fit parameters of the signal, such as the masses, spins and eccentricity.  In this analysis, we are characterizing a signal as being observably eccentric based upon the \ac{SNR} in the different eccentric harmonics.  As shown in \cite{Patterson:2024vbo}, we can use the observed \acp{SNR} to infer the eccentricity.  Thus, we are able to, in a straightforward way, model the impact of noise on the measurability of eccentricity, without performing full parameter estimation analyses.  We neglect the impact of noise on the inferred masses and spins of the binary, as this will have minimal impact on our results.

When combining \ac{SNR} measurements across detectors in quadrature, we must in principle also add the degrees of freedom together, so that the noise contribution is modelled as having $2N$ degrees of freedom for a signal observed in $N$ detectors \citep{Harry:2010fr}.%
\footnote{The number of degrees of freedom for a real matched filter template bank search depends on the detection statistic used. For a coincidence search, the statistic has two degrees for each detector (as in this study), whereas a coherent search has four degrees of freedom irrespective of the number of detectors \cite{Harry:2010fr}. Additionally, many current search techniques use a re-weighted \ac{SNR} to combat non-stationary noise which we do not consider in this work, further affecting the expected \ac{SNR} distribution \cite{Babak:2012zx, Nitz:2017svb}.}
For the sub-leading eccentric harmonics, we can exploit the fact that their amplitude and phase in each detector, relative to the observed $k=0$ harmonic, will be the same.  Thus, we are able to coherently combine the \ac{SNR} in each of the eccentric harmonics across a detector network, with only two unknown parameters: the overall amplitude and phase of the harmonic.  Therefore, each harmonic will be affected by only two noise degrees of freedom.  Furthermore, the relative phases of the $k=1$ and $k=-1$ harmonics are fixed.  Requiring consistency of these phases enables us to remove an additional degree of freedom (see \cite{Patterson:2024vbo} for details) and model the observed to the eccentric \ac{SNR} squared as a non-central chi-squared distribution with three degrees of freedom.


To include these effects of gaussian noise in this study we draw a sample observed value of $\rho_\mathrm{circ}$ and $\rho_\mathrm{ecc}$ from the corresponding distribution for each event, and compare this value to their corresponding threshold condition. This is equivalent to saying that we accept each event based on the probability that a random realization of gaussian noise would give an observed \ac{SNR} that passes the threshold.

We choose a threshold on the eccentric \ac{SNR} of $\rho_{\mathrm{ecc}} \geq 4$.  This is a fairly stringent requirement, corresponding to just a $0.1\%$ chance of falsely classifying a given non-eccentric event as eccentric. However, due to the far greater quantity of non-eccentric events in the population, this can still lead to a significant fraction of events we classified as eccentric to actually be non-eccentric. Let us consider a simplified model where a fraction $x$ of all total events are eccentric, and $(1-x)$ are non-eccentric. We assume that all eccentric events are correctly labeled as eccentric, however non-eccentric events are incorrectly identified as eccentric with a given \ac{FPR}. 

This leads to a \ac{FDR} of
\begin{equation}
\label{eq:FalseEccentric}
\mathrm{FDR} = \frac{(1-x)\cdot\mathrm{FPR}}{x+(1-x)\cdot\mathrm{FPR}}.
\end{equation}
This is the fraction of events identified as eccentric which actually originate from non-eccentric mergers. Since only a small fractions of observed signals are expected to originate from eccentric binaries (i.e. $x \ll 1$), even a very low \ac{FPR} can lead to a non-negligible \acp{FDR}. 
A threshold of $\rho_\mathrm{ecc} \ge 4$, with the eccentric \ac{SNR} distributed as non-central chi-squared with 3 degrees of freedom, corresponds to $\mathrm{FPR} \approx 0.001$. For a universe where $1\%$ ($5\%$) of all detected events are eccentric, this corresponds to an \ac{FDR} of approximately $10\%$ ($2\%$).

\section{Results}\label{sec:results}
\subsection{Observed eccentricity distributions}
\label{subsec:obs_ecc_dist}

Using the standardized eccentricity distribution from \cite{vijaykumar_2024_data_release}, we generate a sample of $10^6$ \acp{BBH} that follow the eccentricity, masses, and aligned spin distributions governed by the \ac{CMC}  catalog \citep{Kremer:2019iul} described in Section~\ref{subsec:ecc_models}. The sources are distributed uniformly in comoving volume and source frame time which is a simple approximation to an astrophysically motivated population. In reality, the source distribution may deviate from strict uniformity due to star formation rate evolution and cosmological effects \citep{Madau:2014bja}. Furthermore, we distribute these \acp{BBH} uniformly in orientation and location in the sky.  For each of the binaries, we calculate the expected \ac{SNR} in the LIGO-Virgo network operating at \ac{O4} sensitivity.  To do so, for each of the \acp{BBH} in the sample, we generate the eccentric \ac{GW} signal and calculate the expected \ac{SNR} for the best matching quasi-circular template, $\rho_{\mathrm{circ}}$ optimized over the chirp mass as described in Sec.~\ref{subsec:detection}.  Any signal with $\rho_{\mathrm{circ}} \ge 10$ is classified as observable in the network.

\begin{figure*}[ht!]
\centering
\includegraphics[width=0.94\textwidth]{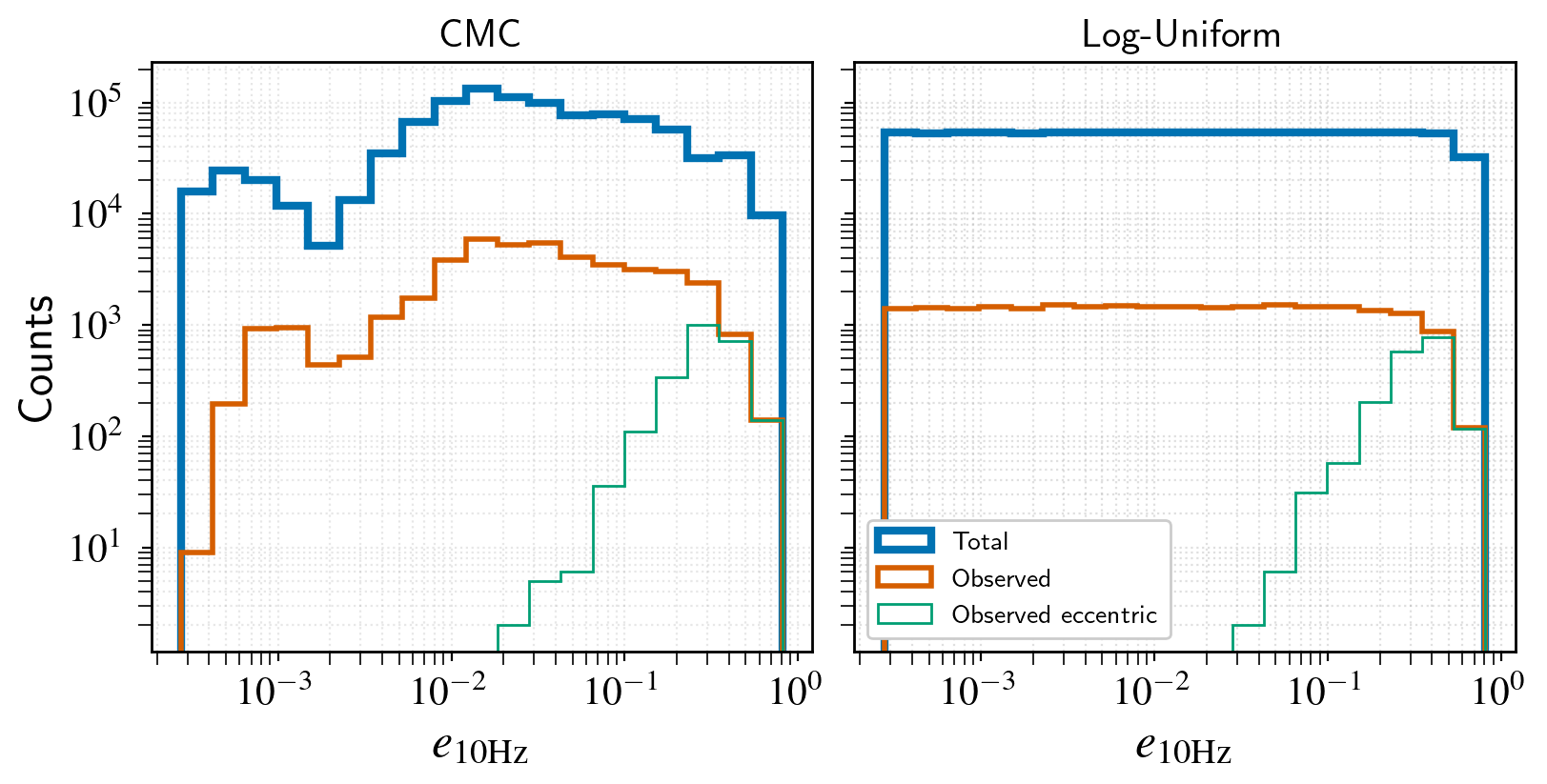}
\caption{The distributions of residual eccentricity at orbit-averaged frequency of $10$Hz for the total population in blue, observed population in orange (assuming matched filtering \ac{SNR} of best matching quasi-circular template is greater than 10), and observed eccentric population in green (assuming eccentric harmonic \ac{SNR} is greater than 4), for \acp{BBH} from \ac{CMC}  simulations (\textit{left}) and the log-uniform model (\textit{right}). In the \ac{O4}  observing scenario of \ac{LVK}, we find $\sim90\%$ of the measurable eccentricities are $\geq 0.2$ for both populations. The \ac{SNR} loss due to filtering eccentric signals against quasi-circular templates affects the detection of \acp{BBH} at larger eccentricities.}
\label{fig:ecc_sim_obs_dist_circ}
\end{figure*}

The left panel of Fig. \ref{fig:ecc_sim_obs_dist_circ} shows the distribution of eccentricities for the \ac{CMC} model, with the eccentricity calculated at an orbit-averaged reference frequency of $10$Hz.  As expected, the distribution of eccentricities for observed binaries is similar to the total population except at high eccentricities where quasi-circular templates are not a good approximation to the eccentric \ac{GW} signal \citep{Martel:1999tm, Brown:2009ng, Huerta:2013qb, Lower:2018seu, Romero-Shaw:2019itr}. This results in a significant loss in the sensitivity of search and hence a loss in the recovered \acp{SNR} making binaries unobservable.   

Next, we perform the eccentric harmonic decomposition, as described in Sec.~\ref{subsec:ecc_harmonics}, for each of the observed \acp{BBH} in the population. We compute the \ac{SNR} in the leading eccentric harmonic ($k=0$) and two subdominant eccentric harmonics ($k=1, -1$). We define the \ac{SNR} due to eccentricity, $\rho_{\mathrm{ecc}}$, to be the quadrature sum of the \acp{SNR} in the $-1$ and $1$ eccentric harmonics, as discussed in Sec.~\ref{subsec:IdentifyEcc}. 
We impose a threshold of $\rho_{\mathrm{ecc}} \geq 4$ for detectability of eccentricity. The distribution of measurable eccentricities also shown in the left panel of Fig. \ref{fig:ecc_sim_obs_dist_circ}. We find that $\sim90\%$ of the \acp{BBH} with measurable eccentricities have values $\gtrsim 0.2$. 

We follow the same steps for the mock \ac{BBH} population with log-uniform eccentricity distribution as described in Section \ref{subsec:ecc_models}. The right panel of Fig.~\ref{fig:ecc_sim_obs_dist_circ} shows the distribution of eccentricities for the log-uniform model. The observation of \acp{BBH} is again suppressed at high eccentricities. The measurable eccentricity distribution is similar to the one in the \ac{CMC}  population, suggesting that the distribution of \acp{BBH} with measurable eccentricity is largely independent of the choice of model chosen for eccentricity and masses.

\subsection{Observed eccentricity against SNR for BBH populations}
\label{sec:ecc_vs_snr_scatter}

\begin{figure*}[ht!]
\centering
\includegraphics[width=0.98\textwidth]{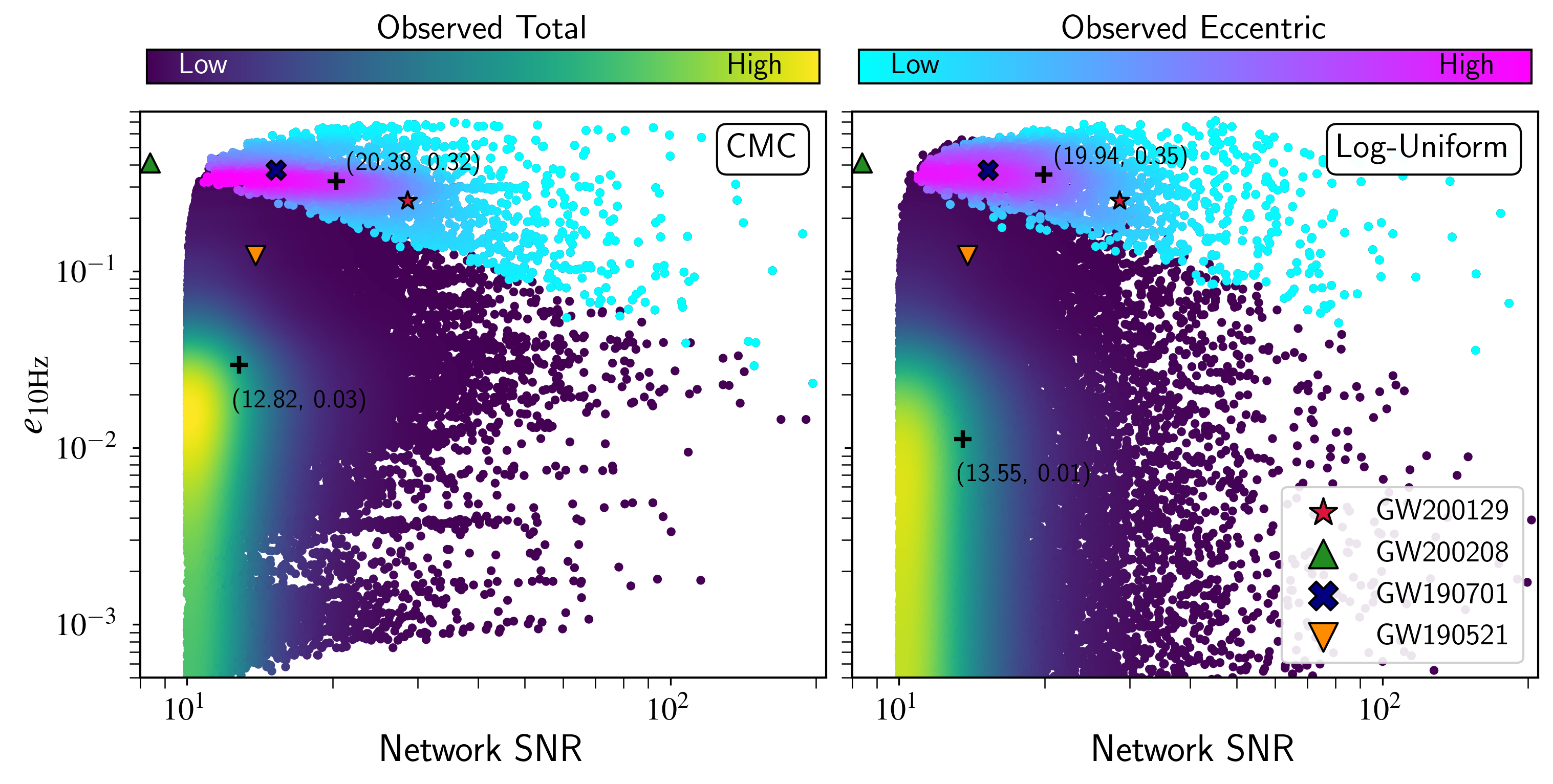}
\caption{The distribution of eccentricity as a function of full expected network \ac{SNR} (assuming eccentric templates) for observed (in dark blue to yellow colormap) and measurably eccentric (in cyan to pink colormap) population of \acp{BBH} using quasi-circular templates. The colormap density here corresponds to the number of binaries. The median eccentricity and \acp{SNR}  (denoted by `\textbf{+}' with median \acp{SNR}  and eccentricities, respectively, quoted in the brackets) for the measurably eccentric population are significantly higher with respect to the observed population of \acp{BBH}. \textit{Right:} Same as left plot but the log uniform eccentricity and PLP mass distributions. The conclusions are largely same as for \ac{CMC} model. This highlights the model independence of the measurably eccentric population distribution of \acp{BBH}.}
\label{fig:ecc_scatter_plot}
\end{figure*}

Investigating the population of \acp{BBH} with measurable orbital eccentricity allows us to predict the most likely eccentricity and \acp{SNR} of the first \ac{BBH} mergers with observable eccentricity. Both higher eccentricity and higher \ac{SNR} contribute to the detectability of a source as eccentric, and we find that the eccentric \ac{SNR} will scale approximately linearly with both. In fact, we find that as a general rule of thumb the eccentric harmonic SNR can be approximated as $\rho_\mathrm{ecc} \approx e_{10}\cdot\rho_\mathrm{total}$.  However, the underlying astrophysical population is dominated by low eccentricity, and the observed population is dominated by low \ac{SNR} (close to the detection threshold) and low eccentricity events. The natural question is therefore whether we expect the detection of eccentricity to be more common for events with high \ac{SNR} and low eccentricity, or low \ac{SNR} and high eccentricity, or whether we require both high \ac{SNR} and high eccentricity to confidently identify the signal as eccentric. 

In Fig.~\ref{fig:ecc_scatter_plot}, we show the residual eccentricity and network \ac{SNR}, computed assuming eccentric templates, for the observable \acp{BBH} (with $\rho_{\mathrm{circ}} \geq 10$) within the \ac{CMC} and log-uniform populations. As highlighted in Sec \ref{subsec:detection}, the \acp{BBH} with high eccentricity and \acp{SNR} close to threshold (top left corner) are not observable due to loss in the sensitivity of the quasi-circular search. The population of observed \acp{BBH} with measurable eccentricity ($\rho_{\mathrm{ecc}} \geq 4$) is overlaid on the observed population. The threshold for detectability of eccentricity roughly follows a straight line, as expected since $\rho_{\mathrm{ecc}}$ scales approximately linearly with both eccentricity and \ac{SNR}. The coloring of the points in Fig.~\ref{fig:ecc_scatter_plot} shows their density. From this, it is clear that the observed population is dominated by signals at, or close to, the detection threshold and is also dominated by signals with low eccentricities $e_{\mathrm{10Hz}} \lesssim 0.05$.  

The measurably eccentric binaries have much larger eccentricities and \acp{SNR} than the bulk of the observed population. For example, the median \ac{SNR} and the residual eccentricity of the observed population are $\sim \cmcObsCircSNRSnrMedian$ ($90\%$ range: $\cmcObsCircSNRSnrLowerBound - \cmcObsCircSNRSnrUpperBound$) and $\cmcObsCircSNREccMedian$ ($90\%$ upper limit: $\cmcObsCircSNREccUpperBound$) whereas the same quantities for the measurably eccentric population are $\cmcObsEccSNRSnrMedian$ ($90\%$ range: $\cmcObsEccSNRSnrLowerBound - \cmcObsEccSNRSnrUpperBound$) and $\cmcObsEccSNREccMedian$ ($90\%$ range: $\cmcObsEccSNREccLowerBound - \cmcObsEccSNREccUpperBound$). This indicates that the average signal which is identified as eccentric should have moderately high network \ac{SNR} and a large eccentricity. However there is clearly a wide range possible for these two parameters. Furthermore, as expected, we see a negative correlation between the minimum detectable value of eccentricity and network \ac{SNR} for the detection of eccentricity, i.e. smaller eccentricities require higher \ac{SNR} to be detected, and vice versa. We also see a small amount of overlap between the populations that are identified as non-eccentric and eccentric due to the dependence of $\rho_\mathrm{ecc}$ on the mass of the binary. These conclusions are robust across both \ac{CMC} (left panel) and log-uniform (right panel) population models used in this work.

\subsection{Comparing eccentric candidates from \ac{GWTC-3}}\label{subsec:ecc_candidates}
\label{subsec:candidates}

There have been several claims of observed eccentricity in \ac{BBH} detections reported in \ac{GWTC-3}. In particular, the most massive \ac{BBH} merger detected in \ac{GWTC-3}, GW190521, was interpreted as eccentric by several studies based on the comparison to numerical relativity waveforms \citep{Gayathri:2020coq} and Bayesian inference with likelihood reweighting \citep{Romero-Shaw:2020thy}, though its features could also be explained by a quasi-circular precessing waveform model \citep{GW190521:2020iuh}. Using a machine learning approach, \citet{Gupte:2024eep} claimed the measurement of eccentricity in GW190701, GW200129, and GW200208\_22 with $\log_{10}\mathrm{Bayes \ factors}$ ranging from $\sim 1.7 - 4.5$ depending on the treatment of noise glitches in the data, although did not find any evidence in support of eccentricity in GW190521. Recently, \citet{Morras:2025xfu} found that an the eccentric hypothesis is preferred to the quasi-circular one for an \ac{NSBH}, GW200105.  However, the evidence for this depends strongly on the choice of priors used for eccentricity distribution in the analysis.

To check the consistency of the measured eccentricities and the SNRs of these various claims against the observed population of \acp{BBH} with measurable eccentricity, we plot the maximum-likelihood estimates for eccentricities and network matched-filter \acp{SNR} for GW190521 from \citet{Romero-Shaw:2020thy}, and GW190701, GW200129, and GW200208\_22 from \citet{Gupte:2024eep} in Fig.~\ref{fig:ecc_scatter_plot}. The eccentricity estimates for these candidate events assume \texttt{SEOBNRv4EHM} waveform model, which leads to an overestimation of the eccentricity compared to \texttt{TEOBResumS-Dali} \citep{Knee:2022hth}. For consistency, we have converted \texttt{SEOBNRv4EHM} eccentricities to the ones corresponding to \texttt{TEOBResumS-Dali} Waveform using match calculations suggested in \citet{Knee:2022hth}. We do not compare the estimates for GW200105 since it is an \ac{NSBH} which we do not consider in this study.

We find that the measured eccentricity and network \acp{SNR} of GW190701 and GW200129 are consistent with the measurably eccentric population found in this work. While GW200208\_22 is at a lower network \ac{SNR} than any of the events in our measurably eccentric population, this is due to the condition we have required for a detection of an event of an \ac{SNR} of 10. Lowering this threshold would therefore likely result in events in our sample having very similar parameters to GW200208\_22. GW190521 on the other hand is claimed to have an eccentricity lower than the majority of \acp{BBH} where we find we can detect eccentricity. If we use maximum-likelihood estimates from \cite{Gayathri:2020coq}, which reports significantly high value of eccentricity $\sim 0.69$ for roughly the same \ac{SNR}, we find that there is not enough power in the quasi-circular \ac{SNR} $\rho_\mathrm{circ}$ in order for the event to be found by quasi-circular search templates. 

\subsection{Impact of noise on the measurement of eccentricity}
\label{subsec:noise}

\begin{figure*}[ht!]
\centering
\includegraphics[width=0.98\textwidth]{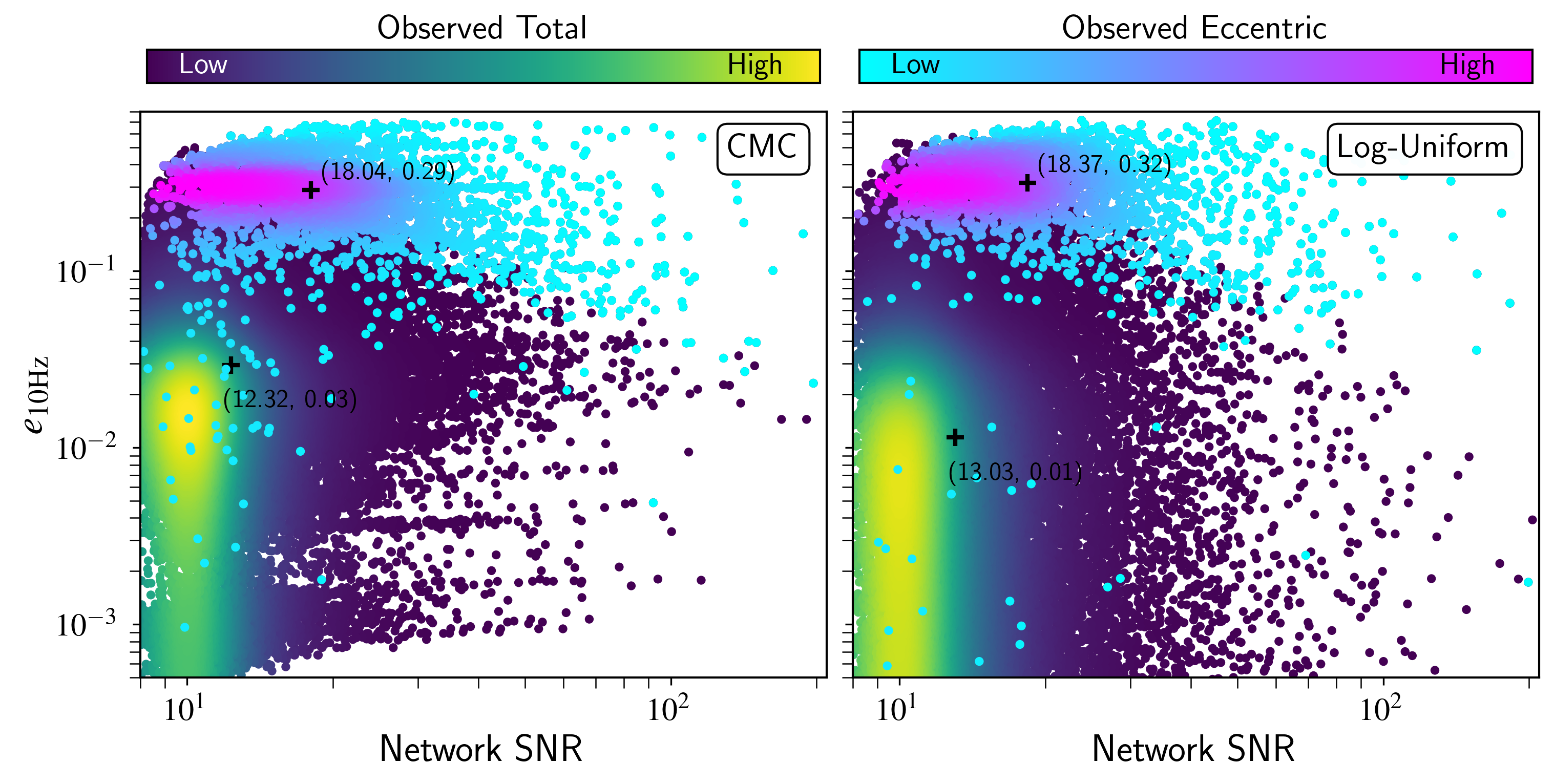}
\caption{Same as Fig. \ref{fig:ecc_scatter_plot} but here the categorization whether the system is observable or measurable eccentric considers the effect of noise. The distribution of the measured eccentric harmonic \ac{SNR} follow non-central chi-square distribution with 3 degrees of freedom and non-centrality parameter being the square of the true eccentric harmonic \ac{SNR}.}
\label{fig:ecc_scatter_plot_noise}
\end{figure*}

We show the effects of including noise in Fig. \ref{fig:ecc_scatter_plot_noise}.  In the plot, we show the true eccentricity and network \ac{SNR} of the signal (in the absence of noise).  We use the observed network \ac{SNR}, $\rho_{\mathrm{circ}}$, and eccentric \ac{SNR}, $\rho_{\mathrm{ecc}}$ with the addition of gaussian noise to determine whether the system would be observed.  The observed distribution now clearly extends to lower \acp{SNR} as the random noise can increase the observed \acp{SNR}.%
\footnote{We have only included events with an expected network \acp{SNR} of at least 8 for computational reasons.  A very small number of events with lower \ac{SNR} may be observed, but this will not affect our conclusions.}
The fact that events with expected \ac{SNR} less than 10 can be observed serves to decrease the median \ac{SNR} of observed events slightly to $\cmcObsCircSNRSnrMedianNoise$. The greater effect, however, is the extension of the ``measurably eccentric'' population to lower eccentricities.  As discussed in Section \ref{subsec:noise}, the $\rho_{\mathrm{ecc}} \ge 4$ threshold corresponds to a $0.1\%$ false acceptance rate.  However, the population is dominated by events with low eccentricity, leading to a not-insignificant fraction of the observably eccentric population arising from the non-eccentric population. For this reason, the median \ac{SNR} and eccentricity of the population are lower than in the noise free case with $\sim \cmcObsEccSNRSnrMedianNoise$ and $\sim \cmcObsEccSNREccMedianNoise$ respectively. It is also clear that even with our stringent condition on $\rho_\mathrm{ecc}$, some events with almost zero residual eccentricity are nevertheless misidentified as eccentric due to their abundance in the observed population.

\subsection{A search using eccentric harmonics}
\label{sec:search}

\begin{figure}[ht!]
\centering
\includegraphics[width=0.45\textwidth]{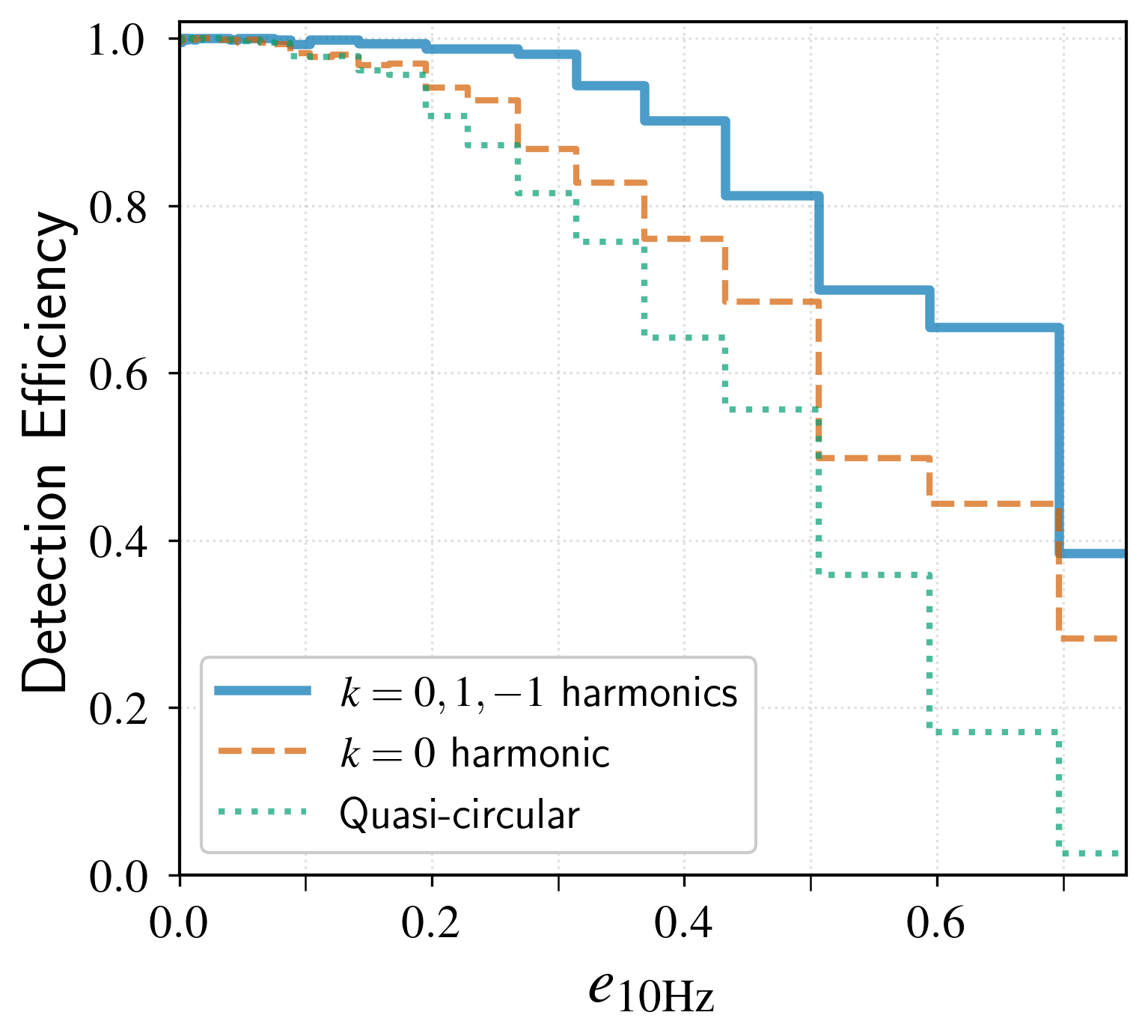}
\caption{The detection efficiency of the quasi-circular (in green), dominant eccentric harmonic $k=0$ (in orange), and two sub-leading eccentric harmonics in adition to dominant harmonic $k=(0, 1, -1)$ (in blue) searches with respect to full eccentric template search.  In all cases, we require a network \ac{SNR} above 10 for detection even though, as discussed in the text, the threshold is likely to be higher for the more complex searches incorporating multiple harmonics or spanning the full space of eccentric signals.  The quasi-circular approach suffers from loss of the sensitivity for \acp{BBH} with $e_{\mathrm{10Hz}} > 0.1$, whereas the $k=0$ harmonic search shows significant improvement over the quasi-circular search for $e_{\mathrm{10Hz}} > 0.2$.}
\label{fig:ecc_detection_efficiency}
\end{figure}

The current modeled \ac{GW} searches used by the \ac{LVK} collaboration do not account for the non-circular orbits of the binaries, which limits the sensitivity of the searches for moderate to highly eccentric \ac{BBH} mergers \citep{Brown:2009ng}. We assess the loss of sensitivity arising from the quasi-circular search templates and compare the expected performance of different eccentric searches in Fig. \ref{fig:ecc_detection_efficiency}. We show the detection efficiency, the fraction of detected binaries using a search compared to the full eccentric search, as a function of residual eccentricity in the population.  

We investigate three possible search methods. The first is a search using quasi-circular templates, as is currently performed in \ac{LVK} analyses.  Second, we evaluate a potential search using the $k=0$ eccentric harmonic.  To perform this search would require the construction of a template-bank which includes orbital eccentricity, in addition to masses and (aligned-)spins of the binary components.  This would inevitably enlarge the number of templates required to cover the search space and therefore increase the false rate at a given \ac{SNR}.  However, since the $k=0$ waveform has a similar structure to a quasi-circular waveform, the existing search architecture should require minimal other changes.  Finally, we consider a search also using the two sub-leading $k=1, -1$ eccentric harmonics in addition to the dominant $k=0$ harmonic.  In this case, the size of the template bank would increase, as would the complexity of the search as three separate waveforms (the $k= -1, 0, 1$ eccentric harmonics) would need to be filtered at each point in parameter space. 

In this initial comparison, we assume an identical detection threshold of $\rho = 10$ across all three searches.  In practice, a larger \ac{SNR} in eccentric searches is likely to be required to obtain the same false alarm rate due to a larger number of template waveforms \citep{Harry:2016ijz}, and a more complex search methodology for the search with sub-dominant harmonics \citep{McIsaac:2023ijd}.  We find that the quasi-circular search loses significant sensitivity to eccentric systems, particularly for $e_{\mathrm{10Hz}} \gtrsim 0.1$, resulting in a loss of $\sim 5\%$ of the observed population, roughly consistent with previous predictions based on fitting factors \citep{Divyajyoti:2023rht}. For eccentricities $\sim 0.3$, the quasi-circular loses almost $20\%$ of the sensitivity of a full eccentric search. As expected, the sensitivity loss is higher for larger eccentricities. Notably, the fraction of non-detected eccentric signals when using quasi-circular templates is robust across the two population models assumed in this work.

Next, when evaluating the performance of the leading $k=0$ eccentric harmonic search, we find that it provides better sensitivity than quasi-circular templates as the eccentric harmonic captures the orbit averaged impact of eccentricity, most notably the faster decay of the orbit. For example, for $e_{\mathrm{10Hz}} \sim 0.3$, the detection efficiency with $k=0$ harmonic is $\sim 5\%$ larger than quasi-circular waveforms, increasing further at higher eccentricities.

\begin{figure}
\centering
\includegraphics[width=0.45\textwidth]{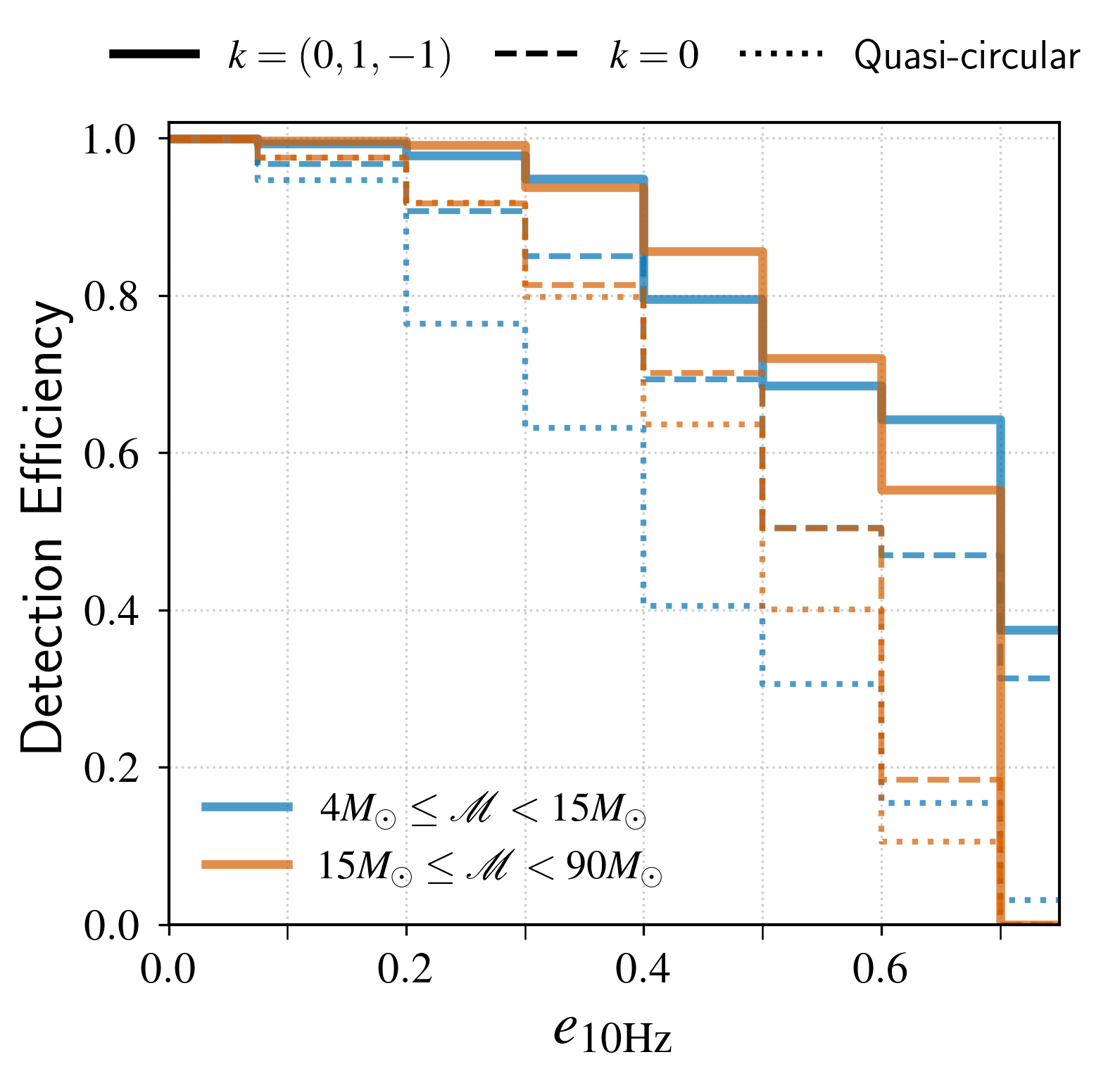}
\caption{The detection efficiency of the quasi-circular (dotted line), dominant eccentric harmonic (dashed line), and three leading eccentric harmonics (solid line) with respect to full eccentric templates in different chirp mass bins. For high masses, the detection efficiency of the dominant harmonic search is comparable to the quasi-circular search whereas three leading eccentric harmonics search is significantly better. On the other hand, for low mass BBHs, the dominant harmonic performs significantly better than quasi-circular, and inclusion of two subleading harmonics further improves the detection efficiency.}
\label{fig:ecc_sim_obs_dist_cum}
\end{figure}

A search using the $k=0$ eccentric harmonic will require a greater number of templates than a quasi-circular search.  However, it is likely to require significantly lower computational cost than a search incorporating the full eccentric waveforms.  There are two reasons for this.  Firstly, eccentric harmonics do not depend on mean anomaly, as can be seen from Eq.~\ref{eq:ecc_harms}, as they average the waveform over this parameter. Thus, rather than adding two additional parameters to search over as in \citet{Dhurkunde:2023qoe, Wang:2025yac}, this would only require extending to eccentricity.  Secondly, the $k=0$ eccentric harmonic waveform varies more slowly with eccentricity than the full waveform, which contains a superposition of multiple harmonics with relative amplitudes depending on the eccentricity \citep{Patterson:2024vbo}.  Furthermore, there is a broad degeneracy between eccentricity and chirp mass.  Therefore, utilizing the dominant eccentric harmonic means that fewer waveforms are required to cover the same range of masses, spins and eccentricities. This motivates the development of eccentric \ac{GW} searches to employ the leading eccentric harmonic to increase search sensitivity without the significant increase in computational cost. 

Finally, we consider the contribution from sub-leading eccentric harmonics $k=1,-1$ to the dominant harmonic search. We find the sensitivity improves significantly compared to full eccentric search and the three leading eccentric harmonics search performs as good as full eccentric search up to eccentricities $\sim 0.3$. This is expected since the three leading eccentric harmonics are sufficient to approximate the full eccentric signals for most of the binaries. Notably, similar searches have been developed for precessing signals and searches incorporating higher \ac{GW} harmonics, which use both the dominant harmonic and one or more sub-dominant contributions \cite{Fairhurst:2019vut, Wadekar:2024zdq, McIsaac:2023ijd}.

In Fig. \ref{fig:ecc_sim_obs_dist_cum}, we investigate the performance of various searches as a function of binary mass.  For high mass binaries, chirp mass $\mathcal{M} \geq 15 M_{\odot}$, the detection efficiency of the dominant eccentric harmonic search is similar to the quasi-circular search while adding subleading eccentric harmonics further improves the sensitivity. However, for low mass ($\mathcal{M} < 15 M_{\odot}$) BBHs, the dominant eccentric harmonic $k=0$ search is significantly better than quasi-circular even for eccentricities as small as $\sim 0.2$. Additionally, the inclusion of two subleading eccentric harmonics further improves the sensitivity of the search.

The sensitivities presented in Figures \ref{fig:ecc_detection_efficiency} and \ref{fig:ecc_sim_obs_dist_cum} illustrate a potential increase in sensitivity from a search for eccentric binaries based upon the harmonic decomposition of the signal.  However, as emphasized above, further investigations are required to demonstrate that such a search would provide an improvement over the standard search for binaries in quasi-circular orbits \citep{GWTC-4}, a fully templated search for eccentric binaries \cite{Dhurkunde:2023qoe} and an unmodelled search for eccentric binaries \citep{LIGOScientific:2023lpe}.  The initial indications suggest that this warrants further study.  A search with either one or three eccentric harmonics clearly provides improved sensitivity to eccentric binaries over the quasi-circular search, for eccentricities $e_{\mathrm{10Hz}} \gtrsim 0.2$ at a fixed \ac{SNR} threshold.  There is likely to be a range of eccentricities where this more than compensates for the increased complexity of the search which would necessitate a somewhat higher detection threshold.  Similarly, searches based on eccentric harmonics are very likely to require fewer templates than the fully eccentric searches, which require filtering as much as one hundred times as many waveforms \citep{Dhurkunde:2023qoe}.  This again suggests there will be a range of masses and eccentricities for which the computational savings and lower detection threshold of the eccentric harmonic search will outweigh any increase in sensitivity from the full eccentric search.

\section{Conclusions}\label{sec:summary}

We have investigated the expected eccentricity distribution of \ac{GW} signals coming from \acp{BBH} in globular clusters for current \ac{GW} detectors and analysis pipelines. Starting from a population of \acp{BBH} informed by \ac{CMC}  simulations of globular clusters \citep{Rodriguez:2017pec, Kremer:2019iul}, we calculate the maximum \ac{SNR} possible for each event by restricting ourselves to matched filtering quasi-circular waveforms, thereby mimicking current matched filtering template banks. Events where this \ac{SNR} is above 10 we judge as being detectable by current \ac{GW} science. Next, we decompose eccentric waveforms into eccentric harmonics at the parameters of each event, following the method introduced by \citet{Patterson:2024vbo}. This enables us to compute an `eccentric harmonic \ac{SNR}' for each event, quantifying the power beyond the dominant eccentric harmonic in each case, and classifying events as measurably eccentric by parameter estimation methods where this \ac{SNR} is greater than 4.

We find that $\sim90\%$ of the \acp{BBH} for which we can detect eccentricity have $e_{10} \geq 0.2$, with the median network \ac{SNR} for these \acp{BBH} significantly higher than the overall observed \ac{BBH} population. We therefore expect the majority of \acp{BBH} for which we can measure eccentricity to be moderately high in both \ac{SNR} and eccentricity. There have been several claims in the literature for detections of eccentricity in events from \ac{GWTC-3} \citep{Gayathri:2020coq, Romero-Shaw:2020thy, Gupte:2024eep}. We find that claims for GW190701 and GW200129 are clear consistent with the measurably eccentric population we find in this work, while GW200208\_22 likely would be consistent if we required a slightly lower quasi-circular \ac{SNR} for event detection. GW190521 however possesses a fairly low eccentricity and we find would likely require a particular noise realization in order to constrain eccentricity away from zero.

Performing this analysis by injected simulated waveforms, applying search algorithms, and employing traditional parameter estimation in order to find the observable population of eccentricity would present serious computational challenges. Our novel approach of examining the \ac{SNR} in the eccentric harmonics enables us to instead make quick and easy approximations of this process, allowing us to apply this analysis to a large simulated population which would not be otherwise feasible. This technique also allows us to cheaply add realizations of stationary, Gaussian noise without the need for re-analysis, as discussed in Section~\ref{subsec:NoiseMethods}.

This work has restricted attention to the ($l=2, m=\pm2$) multipole and aligned spin binaries. We do not expect the first assumption to play a significant role in the results of this study, as higher order multipoles add a small effect to eccentricity, which is already a small effect on the waveform. Ignoring precession, however, could have a much greater effect. Several studies have shown that there is a degeneracy between eccentricity and precession at higher masses, potentially affecting the measurability of eccentricity in this region of parameter space \cite{Romero-Shaw:2022fbf, Xu:2022zza}. We leave the exploration of the impact of this to future work. Finally, although there have been recent suggestions that the \ac{NSBH} event GW200105 may be eccentric \citep{Morras:2025xfu}, we have only considered \acp{BBH} in this study, highlighting another possible future extension of this work.

The restriction of matched filter template bank searches to quasi-circular parameter space has a significant impact on the detection prospects of eccentricity. Events with higher eccentricities, and therefore easier to measure eccentricity, are less likely to be detected in the data due to the \ac{SNR} loss when matched filtering with quasi-circular waveforms. We have shown in this study that conducting a search with the dominant eccentric harmonic is able to find significantly more events than quasi-circular detection methods with a much smaller template bank than a full eccentric search. We also find that including two subleading eccentric harmonics in addition to the dominant eccentric harmonic significantly improves the detection efficiency of eccentric binaries. This suggests that an eccentric search using a template bank of eccentric harmonics may be very effective at detecting eccentric events without prohibitively increasing computational cost, in a similar way to has been applied in the past to higher multipoles \citep{Wadekar:2024zdq} and precession harmonics \citep{Fairhurst:2019vut, McIsaac:2023ijd}. Since the searches explored in this work are a simplified version of standard searches, further investigations would be required to quantify the improvements gained by the eccentric harmonic searches over traditional full eccentric searches. We leave this for future work.




\begin{acknowledgments}
We thank Arif Shaikh for \ac{LVK} Publication \& Presentation review of our manuscript, and Geraint Pratten for his useful comments. We acknowledge help from Aditya Vijaykumar on using the standarized eccentricity data from \ac{CMC}  simulations. We are grateful to Isobel Romero-Shaw for helpful discussions and providing eccentric PE samples for GW190521, and Nihar Gupte for providing eccentric PE samples for GW190701, GW200129, and GW200208\_22. We also thank Mark Hannam for interesting discussions. The authors gratefully acknowledge the computational resources provided by LIGO Laboratory and National Science Foundation grants PHY-0757058 and PHY-0823459 and provided by Cardiff University, and funded by STFC grant ST/I006285/1. MKS  and SF thank to Science and Technology Facilities Council (STFC) for funding through grant ST/Y004272/1. BP thanks the STFC for support through grant ST/Y509152/1. Plots were prepared with Matplotlib with analysis making use of simple-pe \citep{Fairhurst:2023idl}, NumPy \citep{numpy}, pycbc \citep{alex_nitz_2020_3596447}, and Scipy \citep{scipy}.
\end{acknowledgments}








\bibliography{references}{}
\bibliographystyle{aasjournal}



\end{document}